\documentclass[sn-apa]{sn-jnl}% APA Reference Style
\definecolor{PhyRevPink}{RGB}{157,91,179}

\jyear{2021}%

\theoremstyle{thmstyleone}%
\theoremstyle{thmstyletwo}%

\theoremstyle{thmstylethree}%

\usepackage{array}
\usepackage{multirow}

\begin{document}

\title[Pre-service physics teachers' selection of instructional videos for teaching]{(How) Do pre-service  teachers use YouTube features in the selection of instructional videos for physics teaching?}

%%=============================================================%%
%% Prefix	-> \pfx{Dr}
%% GivenName	-> \fnm{Joergen W.}
%% Particle	-> \spfx{van der} -> surname prefix
%% FamilyName	-> \sur{Ploeg}
%% Suffix	-> \sfx{IV}
%% NatureName	-> \tanm{Poet Laureate} -> Title after name
%% Degrees	-> \dgr{MSc, PhD}
%% \author*[1,2]{\pfx{Dr} \fnm{Joergen W.} \spfx{van der} \sur{Ploeg} \sfx{IV} \tanm{Poet Laureate} 
%%                 \dgr{MSc, PhD}}\email{iauthor@gmail.com}
%%=============================================================%%

%\iffalse
\author*[1]{\fnm{Philipp} \sur{Bitzenbauer}}\email{philipp.bitzenbauer@fau.de}

\author[2]{\fnm{Tom} \sur{Teußner}}
%\equalcont{These authors contributed equally to this work.}

\author[3]{\fnm{Joaquin} \sur{Veith}}\email{veith@imai.uni-hildesheim.de}

\author[4]{\fnm{Christoph} \sur{Kulgemeyer}}\email{kulgemeyer@physik.uni-bremen.de}

\affil*[1]{\orgdiv{Professur für Didaktik der Physik}, \orgname{Friedrich-Alexander-Universität Erlangen-Nürnberg}, \orgaddress{\street{Staudtstr. 7}, \postcode{91058} \city{Erlangen}, \country{Germany}}}

\affil[2]{\orgdiv{Institut für Didaktik der Physik}, \orgname{Universität Leipzig}, \orgaddress{\street{Prager Str. 36}, \postcode{04317} \city{Leipzig}, \country{Germany}}}

\affil[3]{\orgdiv{Institut für Mathematik und Angewandte Informatik}, \orgname{Stiftung Universität Hildesheim}, \orgaddress{\street{Samelsonplatz 1}, \postcode{31141} \city{Hildesheim}, \country{Germany}}}

\affil[4]{\orgdiv{Institut für Didaktik der Naturwissenschaften}, \orgname{Universität Bremen}, \orgaddress{\street{Otto-Hahn-Allee 1}, \postcode{28359} \city{Bremen}, \country{Germany}}}

%%==================================%%
%% sample for unstructured abstract %%
%%==================================%%

\abstract{This mixed-methods study examines how pre-service teachers select instructional videos on YouTube for physics teaching. It focuses on the role of surface features that YouTube provides (e.g., likes, views, thumbnails) and the comments underneath the videos in the decision-making process using videos on quantum physics topics as an example. The study consists of two phases: In phase 1, N = 24 (pre-service) physics teachers were randomly assigned to one of three groups, each covering a different quantum topic (entanglement, quantum tunnelling or quantum computing, respectively). From eight options provided, they selected a suitable video for teaching while their eye movements were tracked, and think-aloud data was collected. Phase 2 allowed participants to freely choose one YouTube video on a second quantum topic while thinking aloud. The results reveal a significant emphasis on video thumbnails during selection, with over one-third of the fixation time directed towards them. Think-aloud data confirms the importance of thumbnails in decision-making. A detailed analysis identifies that participants did not rely on (content-related) comments despite they have found to be significantly correlated with the videos' explaining quality. Instead, decisions were influenced by surface features and pragmatic factors such as channel familiarity. Retrospective reflections through a questionnaire support these observations. Building on the existing empirical evidence, a decision tree is proposed to help teachers identify high-quality videos considering duration, likes, comments, and interactions. The decision tree can serve as a hypothesis for future research and needs to be evaluated in terms of how it can help systematize the process of selecting high-quality YouTube videos for science teaching.}


%%\pacs[JEL Classification]{D8, H51}

%%\pacs[MSC Classification]{35A01, 65L10, 65L12, 65L20, 65L70}

\maketitle

\newpage 

\section{Introduction}\label{sec:intro}
Prior research has highlighted the benefits of YouTube for learning \citep[e.g., see][]{rosenthal2018,jackman2019youtube}, including increased engagement, better comprehension, and the flexibility to control the learning experience \citep{Jebe-2019,kay2012,stockwell2015}. Consequently, educators and teachers have recognized the educational value of YouTube, employing it for instructional purposes \citep{manca2016,jung2015}. One of such purposes lies in the integration of YouTube explanatory videos into formal learning environments.

In the last years, there has been extensive science education research on the explaining quality of YouTube explanatory videos: \cite{kulgemeyer2020framework} developed a comprehensive framework for effective explanatory videos, based on guidelines published earlier in the literature 
\citep[e.g., see][]{brame2016,findeisen2019}. For details we refer the reader to the research background section~\ref{sec:RB} of this article. Further studies have investigated the relationship between surface features, such as likes and views, and the explaining quality (i.e., the instructional quality) of YouTube explanatory videos: These studies have revealed that the surface features provided by YouTube may not serve as reliable indicators of the explaining quality of a specific video, while a statistically significant correlation was found between the number of content-related comments and the explaining quality \citep{kulgemeyer2016,bitzenbauer2023}. Based on the above findings, \cite{bitzenbauer2023} emphasize that it is crucial to support teachers ``in selecting videos with high explanation quality from the plethora of (online) resources'' (p. 2) through evidence-based selection criteria.

However, to date, there is a dearth of studies investigating the video selection practices of (pre-service) teachers on YouTube, particularly concerning their decision-making factors. It remains unclear whether teachers rely on YouTube's provided metrics, such as likes, views, or the age of the video, or if they consider the comments section influential. This article addresses this research gap by presenting the findings of a mixed-methods study that explores the decision-making processes of (pre-service) physics teachers when selecting instructional videos on YouTube to be included in learning environments (see methods section~\ref{sec:methods}). The study employs a combination of eye-tracking, think-aloud interviews, and a retrospective questionnaire survey to gain comprehensive insights into the thought processes and strategies employed by pre-service teachers during the video selection process. 

\section{Research background}\label{sec:RB}

Explanatory videos, also referred to as instructional videos, play a vital role in science education research, for example serving as concise introductions and explanations of specific topics of interest \citep{wolf2015erklarstrukturen}. Explanatory videos typically do not exceed 10 minutes in length and have garnered increased attention in both formal and informal learning environments, especially through platforms such as YouTube \citep[e.g., see][]{beautemps2021comprises,pattier2021science}. 

\subsection{Quality criteria of instructional YouTube videos}
\label{sec:qualitycriteria}

Recent scholarly investigations have focused on understanding the factors contributing to the success and popularity of explanatory YouTube videos, particularly in the field of science \citep{beautemps2021comprises,welbourne2016science}. Notably, the video structure has emerged as a crucial determinant in this regard \citep{beautemps2021comprises}.

However, the primary objective of explanatory videos is to support student learning, making the quality of explanations of utmost importance \citep{kulgemeyer2023misconceptions,pekdag2010}. Researchers have explored various frameworks and guidelines to enhance the effectiveness of explanatory videos. For example, \cite{kulgemeyer2020framework} proposed a comprehensive framework for creating effective explanatory videos that aligns with guidelines established earlier by \cite{brame2016} and \cite{findeisen2019}. Furthermore, Kulgemeyer's framework incorporates insights from multimedia learning research and draws upon studies related to instructional explanations conducted by \cite{geelan2012teacher} and \cite{wittwer2008instructional}. The framework encompasses seven factors comprising 14 features that collectively influence the effectiveness of explanatory videos. These factors include video structure, language-level adaptation, minimal digressions, as well as consideration of prior knowledge, misconceptions, and student interest \citep{kulgemeyer2020framework}.

%In his study, 
\cite{kulgemeyer2020framework} empirically tested the effectiveness of the framework by comparing student achievement when exposed to videos developed in accordance with the framework against those that did not strictly adhere to the guidelines. The results demonstrated that students exposed to videos closely aligned with the framework exhibited significantly higher levels of declarative knowledge in post-tests ($d = 0.42$), although no statistically significant difference was observed in post-test scores related to conceptual knowledge.

The correlation between video metrics provided by YouTube, such as the number of views or likes, and the videos' explaining quality has yielded mixed results: \cite{kulgemeyer2016} conducted an exploratory study focusing on instructional YouTube videos on mechanics topics and found that the number of content-related comments posted by users below a specific video was the only variable that correlated significantly with the explaining quality. Conversely, the number of views, likes, and dislikes did not exhibit significant correlations. Similar findings have been brought forth by \cite{kocyigit2019does} who evaluated 53 online videos using the the Global Quality Scale and found the YouTube metrics did not significantly differ across quality groups. \cite{bitzenbauer2023} conducted an additional exploratory study that specifically examined explanatory YouTube videos on quantum topics, namely quantum entanglement and quantum tunnelling. In contrast to earlier findings, the authors observed a small but significant correlation between the number of likes and the quality of explanations in their sample of quantum topic videos ($r = 0.37$, $p < 0.01$).

\subsection{Selection processes of YouTube explanatory videos}
The increasing abundance of low-quality educational content on YouTube has become a matter of concern for researchers \citep[e.g., see][]{tan2013informal,neumann2020evaluating,bohlin2017conceptual} highlighting the crucial role that teachers play in selecting explanatory videos of high quality \citep{jones2011youtube,chtouki2012impact}. This issue is further exacerbated by the reliance on popularity-based rankings in search systems: For instance, \cite{chelaru2012can} observed that the top-10 videos in the YouTube search results received a disproportionately higher number of views, likes, and comments. Additionally, the study by \cite{chavira2021educational} revealed that out of the ten most-viewed videos analyzed in their study, only four were deemed satisfactory in terms of quality.

Despite existing research on YouTube video selection, to the best of our knowledge, no studies have specifically examined the process by which teachers select videos from the list provided by YouTube based on search queries. However, several studies have shed light on user behavior, indicating that individuals often sequentially view the returned videos until they find one that aligns with their needs \citep[e.g., see][]{fyfield2021navigating,tan2011open}. The abundance of choices available on YouTube may contribute to choice overload, making it challenging to identify high-quality content \citep{toffler1984future}. Choice overload describes the phenomenon of increased difficulty in decision-making when faced with a large number of choices \citep{Schwartz-2016}, potentially resulting in decreased motivation to engage with individual options \citep{iyengar2000choice}. 

Against this backdrop, it becomes even more apparent that it is essential to support teachers in the process of selecting explanatory videos for classroom practice. Two main measures have been at the center of the debate so far: 
\begin{enumerate}
    \item Ranked lists of educational channels have been published to ``help Internet users to narrow down their search space by recommending channels'' \cite[][p. 3079]{tadbier2021ranking}. However, ``there is no reason to assume that the extensive offer of ranked lists would not lead to choice overload'' \cite[][p. 3079]{tadbier2021ranking}.
    \item To tackle these challenges, scholars have suggested the utilization of decision-assistance tools like meta-search engines, which employ aggregation techniques \citep{meng2002building,dwork2001rank,haveliwala2002topic} as described by \cite{tadbier2021ranking}.
\end{enumerate}

\section{Research rationale}
\label{sec:RQ}
While we agree that pre-made lists or similar resources might not optimally support (science) teachers in the process of selecting YouTube explanatory videos for classroom practice, we believe that the existing empirical evidence on the explaining quality of YouTube explanatory videos might indeed be useful to facilitate the systematization of teachers' decision-making process. As sketched in section~\ref{sec:qualitycriteria}, several studies have brought forth hints for instructional quality of online explanatory videos and might, hence, provide evidence in this regard:

\begin{itemize}
    \item \cite{bitzenbauer2023} found a statistically significant correlation ($r = 0.46$, $p < 0.001$) between explaining quality and the number of content-related comments in YouTube videos on quantum topics. Similarly, \cite{kulgemeyer2016} reported a significant correlation ($r = 0.38$, $p < 0.01$) between explaining quality and the number of relevant comments for videos on Newton's third law and Kepler's laws, respectively.
    \item \cite{bitzenbauer2023} discovered a significant correlation ($r = 0.37$, $p < 0.01$) between the number of likes and video explaining quality.
    \item It is important to exercise caution when considering additional metrics provided by YouTube, such as the number of views, as previous research has not found stable correlations with explaining quality.
\end{itemize}

\begin{figure}[H]
    \centering
    \includegraphics[width=.8\textwidth]{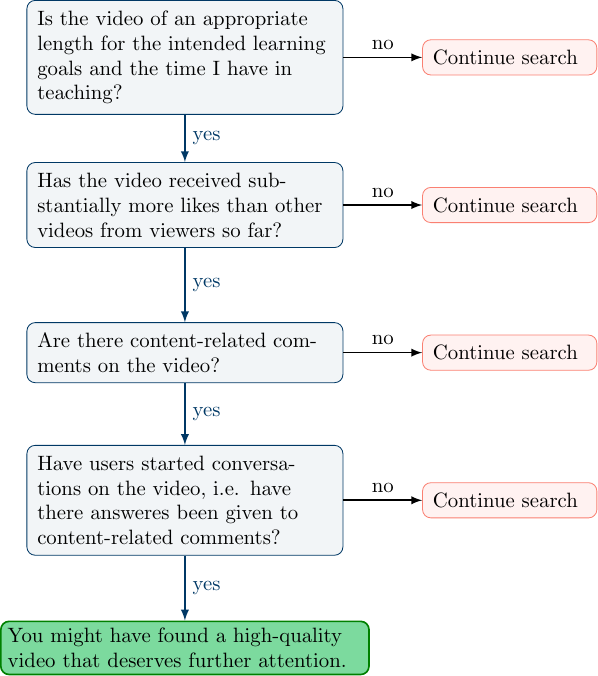}
    \caption{Decision tree to support (pre-service) teachers selection process  when searching explanatory videos suitable for physics teaching as hypothesized based on the current state of the literature. The proposed order is not to be considered strict.}
    \label{fig:flow-chart} 
\end{figure}

Of course, reviewing comments under each video in search of high-quality explanatory videos is practically unfeasible and time-consuming. Moreover, based on the available evidence, it is challenging to determine a quantitative threshold indicating an adequate number of content-related comments or likes. Nonetheless, considering the current state of research, it appears feasible to explore ideal (i.e., efficient) decision-making processes by systematically analyzing the order in which the different criteria can be employed by teachers: As a starting point for future research aimed at supporting teachers in their decision-making processes when selecting YouTube explanatory videos for science teaching, we propose the decision tree presented in Figure~\ref{fig:flow-chart} as a representation summarizing the hints of instructional quality of explanatory videos according to the current state of research described above.

This decision tree suggests teachers to first ask quick initial questions during their search process such as whether a given video has an appropriate duration for classroom use or whether it has already received user likes. If both of these surface-level criteria are positive, it recommends teachers to explore the comments section. The presence of not only superficial but also content- or video-related comments indicates cognitive stimulation of viewers, as ``videos that accumulate plenty of those relevant comments are more successful in catching viewers’ attention as these videos might use either a more stimulating explanation or the explanation delivered is considered as a starting point for further learning progress'' \citep[][p. 12]{kulgemeyer2016}. Moreover, if there are even interactions among users, including responses to content-related comments, this may provide additional evidence of a high-quality video. Finally, teachers are then encouraged to assess the instructional quality of the video by viewing it themselves.

The suggested decision tree serves as a hypothesis for future studies examining teachers’ selection processes of instructional videos for science teaching as stated above. Due to the current lack of empirical studies on (pre-service) physics teachers' decision-making processes when choosing YouTube explanatory videos for the physics classroom, we have two main objectives in this article: First, we aim to explore how pre-service teachers utilize YouTube metrics when selecting instructional videos for physics teaching. Secondly, we intend to compare their selection and decision-making processes with the procedure recommended by the existing literature, as represented through the decision tree in this article.\\
\\
Hence, with our research we aim to approach the following research question: How -- if at all -- do prospective teachers use the features and comments sections provided by YouTube when selecting YouTube explanatory videos for teaching purposes? 

We decided to address this question in the context of quantum physics YouTube videos because (a) as mentioned above, there have been related studies published previously we can build upon with our findings and (b) quantum physics deals with difficult-to-grasp topics and different visualizations or explanations are common to describe the same phenomena due to their abstract nature. Thus, a highly varying degree of explaining quality is to be expected when exploring explanatory videos for topics like quantum tunneling or quantum entanglement. 

\section{Methods}\label{sec:methods}

\subsection{Study design}
\begin{figure}[H]
    \centering
    \includegraphics[width=.85\textwidth]{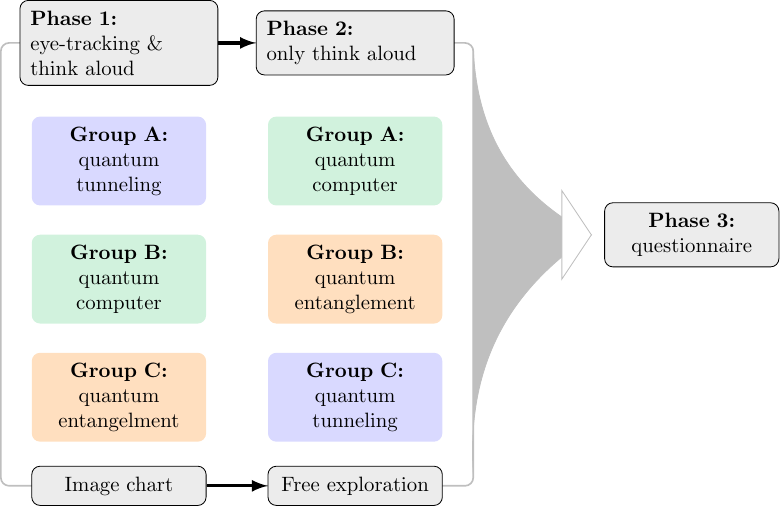} 
    \caption{Study design visualized using a flowchart. The different topics covered in the first two phases are color-coded and, as indicated, were switched among the groups A to C between both phases.}
    \label{fig:design} 
\end{figure}

We investigated our research question by conducting a mixed methods study comprised of eye-tracking, think aloud interviews and a concluding questionnaire. The selection of precisely these methods as well as their interrelation will be explained more thoroughly in the following subsections~\ref{sec:m1}-\ref{sec:m3}. The mixed methods study consisted of three phases (P1 to P3, for an overview see Figure~\ref{fig:design}):
\begin{enumerate}[\hspace{16pt}]
    \item[\bf P1:] In the first phase, the participants were presented with a pre-constructed image chart showing eight different YouTube video suggestions for a specific topic via the original surface features provided by YouTube (e.g., thumbnail, length, title, views, upload date, channel name, number of subscribers). As additional information, we added the corresponding number of likes the videos have already received (cf. Figure~\ref{fig:heat}). The participants were then asked to select one of the offered explanatory videos that they deemed suitable for use with learners without prior knowledge in physics teaching on this topic. The videos displayed in the chart, however, could not be opened or watched and, instead, the selection had to be based solely on the provided features. In addition to tracking the eye-movements during the selection process (cf. Section~\ref{sec:m1}), the participants were prompted to voice their thoughts at all times in the sense of a think aloud interview (cf. Section~\ref{sec:m2}).
    \item[\bf P2:] In the second phase, the image chart was removed and the participants were now allowed to freely explore a previously opened YouTube search tab concerned with a second specific quantum topic. Again, the task was to select one of the videos suggested by YouTube for a hypothetical physics classroom lesson covering the specific topic with learners without prior knowledge. In contrast to the first phase, the participants' eye-movement was no longer tracked but they were now also allowed to open and watch the videos as well as scroll through the comment section. This way, the selection process could place a more pronounced focus on content related reasons to substantiate the rather superficially guided process in phase 1. Similar to phase 1, all thought processes had to be conveyed verbally at all times.
    \item[\bf P3:] After the initial combination of eye-tracking and think aloud, the study concluded with a questionnaire given to students in retrospective which asked the study participants to reflect on the importance of the different surface features provided by YouTube in their selection processes (cf. Section~\ref{sec:m3}). 
\end{enumerate}
To ensure that the results of our study are not directly dependent on (and hence, restricted to) a specific (quantum) topic covered throughout the phases, we randomly assigned each study participant to one of three different groups A, B or C prior to starting the data collection. Each participant took part in the study individually and was -- depending on the group assignment -- given the task of selecting explanatory YouTube videos on two different quantum topics (namely, either quantum tunneling, quantum entanglement, or quantum computing) in study phases 1 and 2 as described above. A subsequent one-way ANOVA comparing the different eye-tracking metrics investigated in this study (cf. Section~\ref{sec:data-analysis}) across the three study groups revealed no statistically significant differences among the groups. This indicates that our results are not directly linked to a specific topic but it is sensible to analyze the data collected in this study across the groups -- we took advantage of this observation in our study, as we analyzed the data from all 24 participants simultaneously, as if they had been collected under exactly the same conditions. An overview of our study design is presented in Figure~\ref{fig:design}.

\subsection{Data collection}
\subsubsection{Eye-Tracking}\label{sec:m1}
Eye-tracking data was collected in study phase 1 using a stationary head-free eye-tracking system from Tobii (Tobii Pro Fusion) alongside their respective software (Tobii Pro Lab). The eye tracker operates at a sampling frequency of 60 Hz and a nominal spatial accuracy of $<0.3^\circ$ visual angle. The stimuli were presented on a 24-inch computer screen (1920 $\times$ 1080 pixel resolution and 60 Hz frame rate). Prior to the study, a nine point calibration process was utilized to ensure accurate eye-tracking and the participants were introduced to the basics of eye-tracking. The instructor verified the agreement between the measured gaze positions and the actual points on the screen. If the calibration results were not deemed satisfactory, the calibration was repeated. In instances where the eye tracker failed to detect sufficient calibration data, participants were repositioned in front of the eye tracker. Additionally, potential factors that could have interfered with pupil detection were examined. On average, the distance between each participant and the tracker was 60 cm.

\subsubsection{Think aloud}\label{sec:m2}
Since previous research has identified the need to complement eye-tracking data with additional verbal data we amended the eye-tracking results by incorporating think aloud interviews into our study design \citep{chien2015learning,smith2010eye}. During think aloud interviews, ``participants think out aloud while performing a given task, or recall thoughts immediately following completion of that task'' \cite[p. 514]{Eccles-17}. The participants' verbalizations were recorded as well as transcribed and subsequently subjected to further analysis (cf. Section~\ref{sec:data-analysis}). The goal of this method lies in uncovering cognitive processes that are not as accessible with the other methods used~\citep{Rios-19}. Thus, even though it might interfere with the study objective due to the verbalizations resulting in an overall higher cognitive demand, think aloud studies are often used as an introspective annex~\citep{Sasaki-13,McKay-09}. We leveraged this method by asking the participants to articulate their thought processes at any point in time in both, study phases 1 (image chart) and 2 (free exploration). To use thinking aloud as effectively as possible in the study, the participants were provided with an instruction on thinking aloud, which was formulated following \cite{mackensen2004forderung} to ensure a standardized procedure \citep{sandmann2014lautes}. Some of the cues given to the participants were: (1) Speak your thoughts aloud, (2) There should be no pauses in speaking, so verbalize your thoughts without pauses, (3) Do not organize your thoughts before speaking, imagine you are alone in the room, (4) Thinking aloud may be a bit unfamiliar. Therefore, you will be supported and repeatedly prompted to express your thoughts.

The additional verbal data obtained from those interviews provides further insights into the cognitive processes, motivations, and decision-making that underlie the observed eye movements, offering a more complete picture of participants' experiences and interpretations~\cite[cf.][]{bruckner2020epistemic}. Furthermore, eye-tracking data alone can identify moments of attention shifts or fixations on specific elements, but it may not explain the reasons for these shifts. Since our study addresses selection processes based on visual stimuli, supplemental verbal data can clarify whether a shift in gaze was triggered by interest, confusion, or any other factors, shedding further light on the nature of the participants' attentional patterns. 

\subsubsection{Questionnaire}\label{sec:m3}
To further enhance cross-validity we concluded our study with a final questionnaire in phase 3. Here, participants were asked to rate whether they considered the different (surface) features provided by YouTube important to their decision-making processes on a four-point rating scale (strongly disagree, disagree, agree, strongly agree). The addressed features were number of views, likes, comments and subscriptions as well as thumbnail, channel, video title, video length, video description, upload date, order determined by YouTube's search algorithm and specific comments. One the one hand, these ratings allow to establish a ranking among all surface features in terms of their importance. On the other hand, the retrospective view obtained from the concluding questionnaire contrasts the introspective view from phases 1 and 2, enabling a triangulation with the findings from both the eye-tracking and the think aloud interviews.

\subsection{Sample}
A total of $N=24$ German pre-service physics teachers ($9$ female, $15$ male) who were at least in their second year of study participated in this research. None of the participants relied on strong glasses or contact lenses (diopter $>1$). The participation in our study was voluntary, not financially recompensed and informed consent was obtained from all participants.

\subsection{Data analysis}\label{sec:data-analysis}
\begin{figure}[H]
    \centering
    \includegraphics[width=\textwidth]{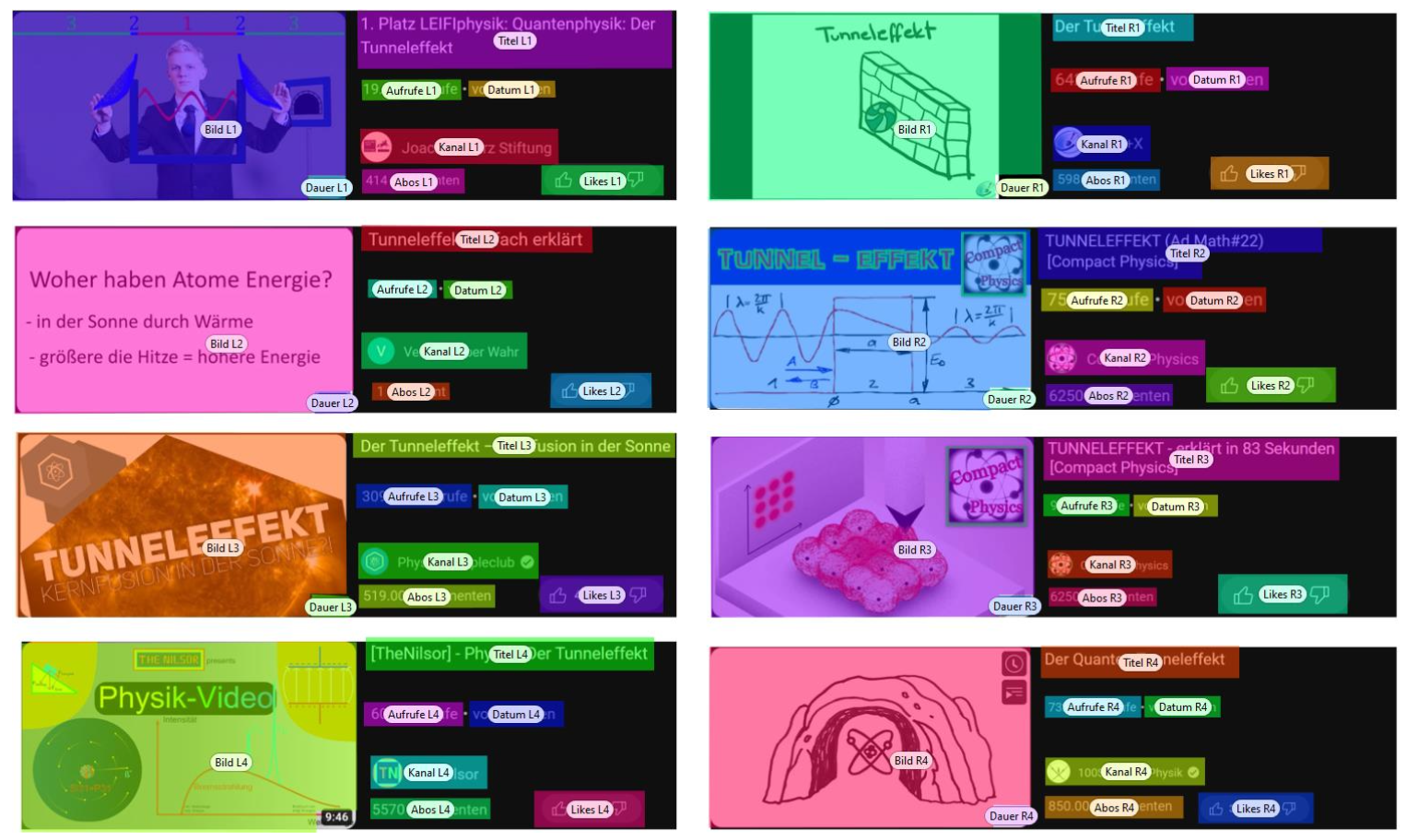} 
    \caption{The AOIs were defined covering all the surface features given for the eight video options shown to the participants. The eye-tracking metrics regarding the related AOIs were summarized using aggregated tags as described in the body text.}
    \label{fig:AOI} 
\end{figure}

The eye-tracking data were evaluated in terms of three metrics that reflect attention allocation and cognitive demand: First, we analysed the \textit{total fixation duration}, which can be described as ``the total duration of all fixations on a specific
stimulus''~\cite[p. 1]{Laan-15}. High values of this metric indicate a more pronounced focus on certain areas~\citep{hahn2022eye}, thus it is the commonly reported measure of visual attention~\citep{Goyal-15}. Second, we investigated the metric \textit{fixation counts} that often accompanies the total fixation duration as a measure of attention allocation~\cite[cf.][]{Wang-14,Just-76}.
Lastly, we analysed the \textit{mean fixation duration}, which is often interpreted as a measure of cognitive processing demand~\citep{Negi-20}. Consequently, higher values of mean fixation duration indicate a ``higher cognitive effort to process information''~\cite[p. 10]{hahn2022eye}. The areas of interest (AOIs) required for quantitative analysis were matched with the surface features provided by YouTube (cf. Section~\ref{sec:m1}) as is indicated in Figure~\ref{fig:AOI}. However, for the data analysis, the individual AOIs for each of the proposed videos shown in the image chart in phase 1 were not considered: Instead, so-called aggregated tags were created that combine the eye-tracking metrics for several related AOIs (e.g., all Like-AOIs). For instance, we combined all the data from the Like-AOIs using an aggregated tag ``Likes''. This allowed us to analyze how frequently and for how long participants viewed the number of likes across different video options.

For the subsequent think aloud interviews we conducted a qualitative content analysis. To this end, we (a) associated the participants' verbal expressions with the corresponding surface features provided by YouTube and (b) categorized their decisions for or against each video. The categories for these decisions were developed based on both inductive and deductive procedures~\citep{Forman-07}. An overview of all categories and their descriptions is provided in the appendix (cf. Table~\ref{tab:cat-system}). The category system was applied by two independent raters and dissenting judgements were resolved through discussion. The interview data were analysed threefold: First, we calculated the relative speaking time allocated to each (surface) feature and visually displayed the resulting share of each feature in a bar chart. Second, we visualized the temporal trajectory of each interview through the various categories and plotted them alongside a common axis, resulting in a temporal topography graphic for each of the two phases. Lastly, we counted the most frequently used arguments among the participants' reasonings and how often they lead to a decision for or against a video. This insight were used to provide an overview of the (surface) features provided by YouTube for each video that are most influential during the decision-making process of pre-service physics teachers.\\
\\
The responses of the concluding questionnaire were summarized using a diverging stacked bar chart are constructed from the participants' responses by aligning the bars from a stacked bar chart relative to the scale's centre~\citep{Robbins-11}. In addition, each of the response options was color coded and equated with a number (strongly disagree $\widehat{=} \ -2$, disagree $\widehat{=} \ -1$, agree $\widehat{=} \ 1$, agree completely $\widehat{=} \ 2$) so that a mean agreement value for each surface feature could be calculated, resulting in a ranking among all surface features provided by YouTube~\cite[cf.][]{Veith-22}.

\section{Results}\label{sec:results}
In the following, we present the results of our study, separated by methodology. First, we provide an overview of the assessed eye-tracking metrics (cf. Section~\ref{sec:results-1}) and second, we enrich those findings with the results of the think-aloud data (cf. Section~\ref{sec:results-2}) as well as the subsequent questionnaire study (cf. Section~\ref{sec:results-3}).
\subsection{Eye-tracking results}\label{sec:results-1}
\begin{figure}[H]
    \centering
    \includegraphics[width=\textwidth]{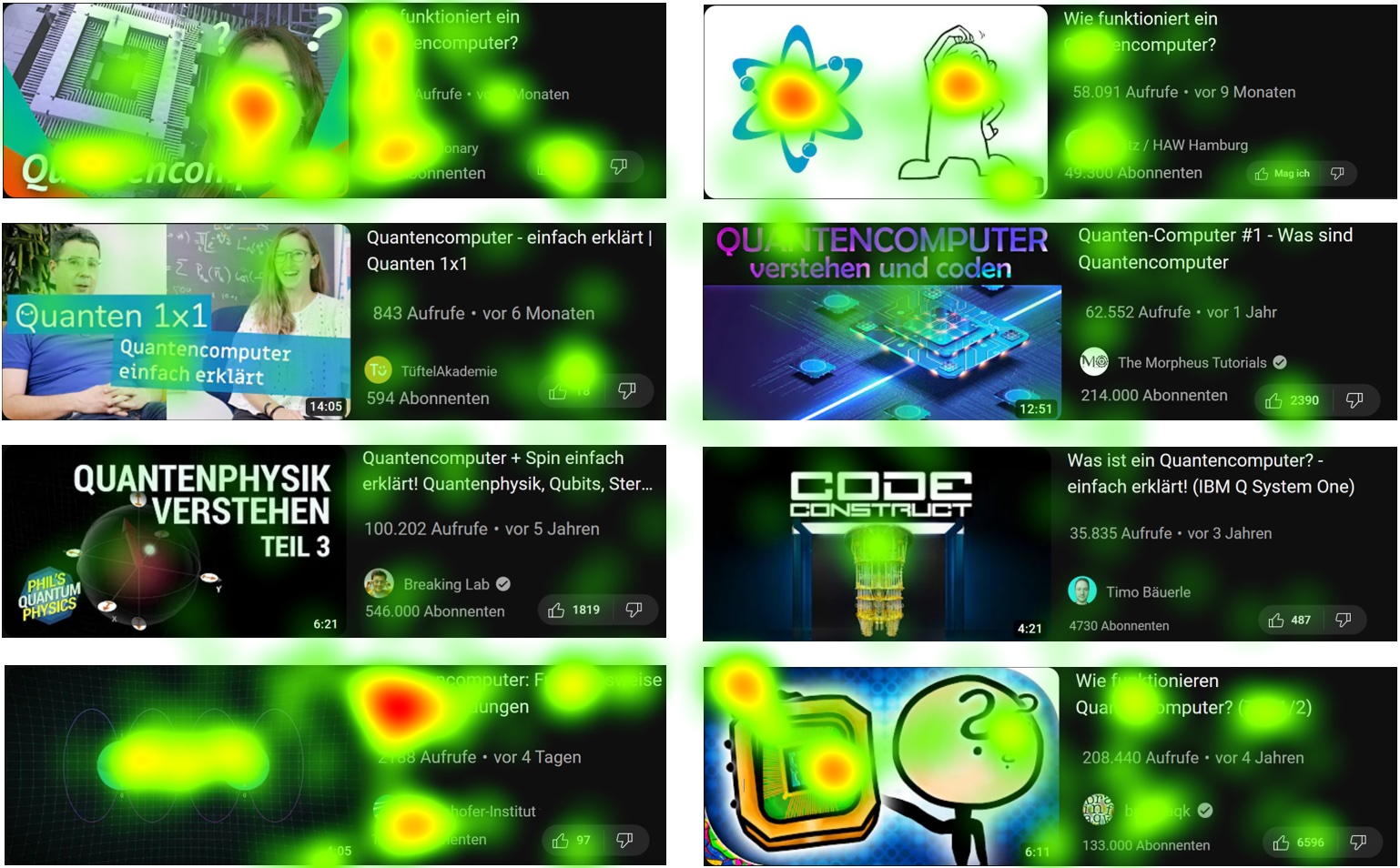}
    \caption{Exemplary heat map created from the eye-tracking data from one of the study participants in phase 1 of our study. Qualitatively, participants' spots of attention are represented through red color. In the following, we provide a quantitative analysis of the data underlying the participants' eye movements.}
    \label{fig:heat} 
\end{figure}

The eye-tracking data were collected in study phase 1 where participants were presented with a carefully constructed chart containing eight search results from YouTube (for details see methods section~\ref{sec:methods}). These options were specifically chosen to exhibit a range of surface features. The participants' task was to determine which of the eight explanatory videos would be suitable for inclusion in a learning environment related to the topic being investigated. Figure~\ref{fig:heat} presents an illustrative heat map generated from the eye-tracking data obtained from one of the participants in the study. The heat map provides visual information regarding the areas that received the highest attention during the task. 

\begin{figure}[H]
    \centering
    \includegraphics[width=\textwidth]{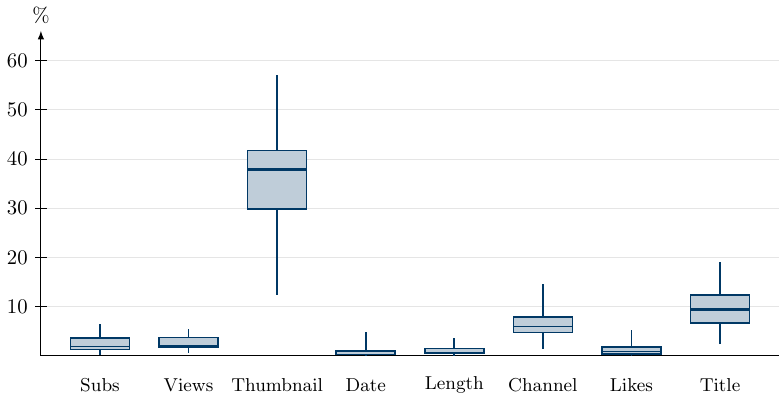}
    \caption{Boxplots for the total fixation duration for each area of interest. The whiskers indicate $1.5\times \text{IQR}$, where IQR is the interquartile range.}
    \label{fig:tot-duration} 
\end{figure}

Table~\ref{tab:tot-duration} provides the descriptive statistics for the metric \textit{total fixation duration} for each area of interest. With a mean percentage of 35.39\% of the total fixation duration, the Thumbnail was by far the most compelling AOI and the only one with a share above 10\%. While the AOIs Title (9.58\%) and Channel (6.17\%) were also able to captivate the participants' attention to some extent, the remaining AOIs played a seemingly negligible role during the selection process. This discrepancy is visualized via boxplots in Figure~\ref{fig:tot-duration}.

\begin{table}[h]
\begin{center}
\begin{minipage}{\textwidth}
\caption{Descriptive statistics for the total fixation duration (in \%) for each area of interest. In addition to the minimum, maximum and mean values, we report the lower ($Q_1$), middle ($Q_2$) and upper ($Q_3$) quartiles.}\label{tab:tot-duration}
\begin{tabular}{c|cccccccc}
 & Subs & Views & Thumbnail & Date & Length & Channel & Likes & Title\tabularnewline
\hline 
\hline 
Max & 6.48 & 5.48 & 57.03 & 4.85 & 3.66 & 14.53 & 5.24 & 19.10\tabularnewline
%\hline 
Min & 0.19 & 0.57 & 12.38 & 0.00 & 0.00 & 1.47 & 0.00 & 2.39\tabularnewline
%\hline 
$Q_{1}$ & 1.28 & 1.73 & 29.84 & 0.29 & 0.46 & 4.73 & 0.32 & 6.69\tabularnewline
%\hline 
$Q_{2}$ & 1.89 & 2.20 & 37.39 & 0.56 & 0.71 & 5.90 & 0.95 & 9.45\tabularnewline
%\hline 
$Q_{3}$ & 3.61 & 3.72 & 41.73 & 0.94 & 1.50 & 7.84 & 1.79 & 12.39\tabularnewline
%\hline 
Mean & 2.55 & 2.47 & 35.39 & 0.78 & 1.08 & 6.17 & 1.56 & 9.58\tabularnewline
%\hline 
\end{tabular}
\end{minipage}
\end{center}
\end{table}

\begin{figure}[H]
    \centering
    \includegraphics[width=\textwidth]{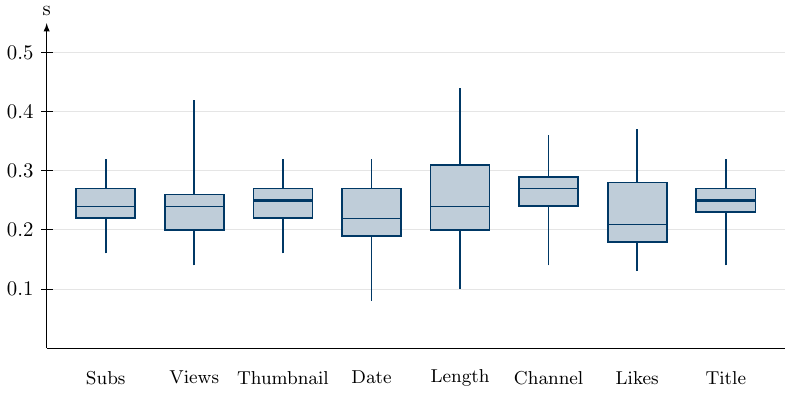}
    \caption{Boxplots for the mean fixation duration for each area of interest. The whiskers indicate $1.5\times \text{IQR}$, where IQR is the interquartile range.}
    \label{fig:avg-duration} 
\end{figure}

Descriptive statistics for the metric \textit{mean fixation duration} are provided in Table~\ref{tab:avg-duration}. The data in this regard paint a different picture by taking similar values for each AOI. The mean fixation durations average between 176~ms (Subs) and 240~ms (Date) and, thus, lie in the typical range of 100-600~ms reported by~\cite{hahn2022eye}. The boxplots illustrating these data substantiate this relationship -- the middle 75\% of data can be located between 200ms and 300ms across almost all AOIs (cf. Figure~\ref{fig:avg-duration}). 

\begin{table}[h]
\begin{center}
\begin{minipage}{\textwidth}
\caption{Descriptive statistics for the mean fixation duration (in seconds) for each area of interest. In addition to the minimum, maximum and mean values, we report the lower ($Q_1$), middle ($Q_2$) and upper ($Q_3$) quartiles.}\label{tab:avg-duration}
\begin{tabular}{c|cccccccc}
 & Subs & Views & Thumbnail & Date & Length & Channel & Likes & Title\tabularnewline
\hline 
\hline 
Max & 0.31 & 0.41 & 0.31 & 0.31 & 0.44 & 0.35 & 0.37 & 0.32\tabularnewline
Min & 0.16 & 0.14 & 0.16 & 0.08 & 0.09 & 0.13 & 0.13 & 0.13\tabularnewline
$Q_{1}$ & 0.21 & 0.20 & 0.21 & 0.18 & 0.20 & 0.24 & 0.17 & 0.23\tabularnewline
$Q_{2}$ & 0.23 & 0.23 & 0.25 & 0.22 & 0.23 & 0.26 & 0.21 & 0.25\tabularnewline
$Q_{3}$ & 0.27 & 0.26 & 0.26 & 0.26 & 0.31 & 0.28 & 0.27 & 0.27\tabularnewline
Mean & 0.17 & 0.19 & 0.21 & 0.24 & 0.18 & 0.23 & 0.21 & 0.23\tabularnewline
\end{tabular}
\end{minipage}
\end{center}
\end{table}

\begin{figure}[H]
    \centering
    \includegraphics[width=\textwidth]{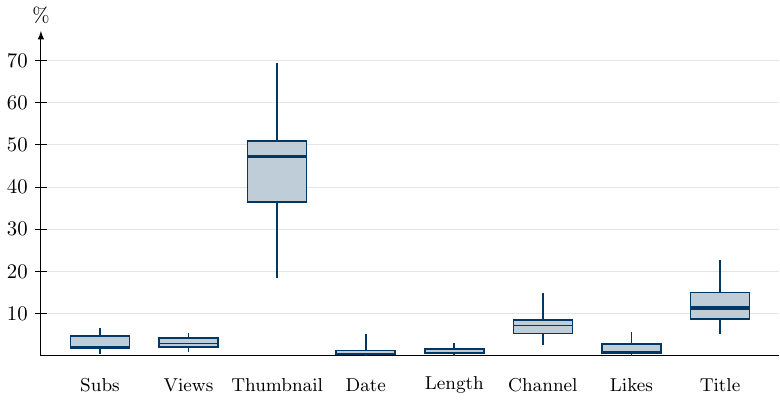}
    \caption{Boxplots for the fixation counts for each area of interest. The whiskers indicate $1.5\times \text{IQR}$, where IQR is the interquartile range.}
    \label{fig:fix-count} 
\end{figure}

Lastly, we investigated the metric \textit{fixation counts} for each AOI. With there being striking differences in the total fixation durations of each AOI but similar values for mean fixation duration, it becomes apparent that the number of fixations must vary in a manner similar to the total fixation duration. The data provided Table~\ref{tab:fix-count} paint a coherent picture in this regard. Again, with a mean percentage of 45.59\% of the total number of fixations, there is a predominant focus on the AOI Thumbnail, with Title (11.60\%) and Channel (7.12\%) being second and third, respectively. Consequently, the boxplots for the metrics total fixation duration and fixation count are somewhat congruent (cf. Figure~\ref{fig:fix-count}).

\begin{table}[h]
\begin{center}
\begin{minipage}{\textwidth}
\caption{Descriptive statistics for the fixation count (in \%) for each area of interest. In addition to the minimum, maximum and mean values, we report the lower ($Q_1$), middle ($Q_2$) and upper ($Q_3$) quartiles.}\label{tab:fix-count}
\begin{tabular}{c|cccccccc}
 & Subs & Views & Thumbnail & Date & Length & Channel & Likes & Title\tabularnewline
\hline 
\hline 
Max & 6.50 & 5.34 & 69.30 & 5.20 & 2.93 & 14.93 & 5.60 & 22.63\tabularnewline
Min & 0.37 & 0.94 & 18.34 & 0.00 & 0.00 & 2.60 & 0.00 & 5.14\tabularnewline
$Q_{1}$ & 1.83 & 2.14 & 36.39 & 0.46 & 0.59 & 5.25 & 0.64 & 8.69\tabularnewline
$Q_{2}$ & 2.29 & 3.04 & 47.29 & 0.93 & 1.04 & 6.55 & 1.00 & 11.29\tabularnewline
$Q_{3}$ & 4.67 & 4.21 & 50.84 & 1.25 & 1.59 & 8.41 & 2.83 & 14.96\tabularnewline
Mean & 3.04 & 3.11 & 45.59 & 1.06 & 1.12 & 7.12 & 1.82 & 11.60\tabularnewline
\end{tabular}
\end{minipage}
\end{center}
\end{table}

\subsection{Think-aloud interview results}\label{sec:results-2}
\begin{figure}[H]
    \centering
    \includegraphics[width=\textwidth]{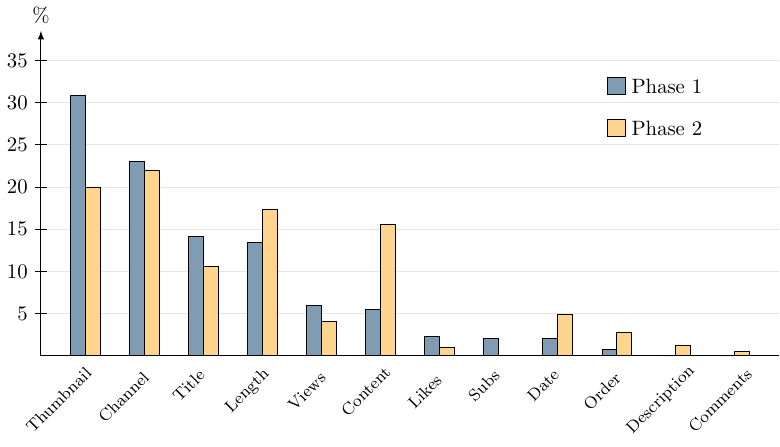}
    \caption{Bar chart visualizing the relative proportion (of total speaking time) of each surface feature in participants' utterances in both study phases. Phase 1 indicates the allocated speaking time regarding the pre-constructed image chart, while phase 2 (yellow) indicates the allocated speaking time during free exploration (cf. Section~\ref{sec:methods}).}
    \label{fig:perc-talk} 
\end{figure}

To analyze the selection process more thoroughly, the results from the eye-tracking study are now complemented with data from a think-aloud study, as described in Section~\ref{sec:methods}. Figure~\ref{fig:perc-talk} offers an initial comprehensive overview of the content aspects addressed in the argumentation provided by the study participants: It shows the relative proportion (of total speaking time) of each surface feature in participants' utterances in both study phases. In phase 1, the most pronounced focus was placed on the thumbnail with 30.9\%, followed by Channel and Title with 23.0\% and 14.2\%, respectively. Regarding the free exploration in phase 2, however, the data convey a more differentiated impression: Here, the surface feature channel emerges on the top with 22.0\%, with the thumbnail (20.0\%) and length (17.4\%) taking second and third places. In addition, the surface features likes, subs, date, order, description and comments are almost negligible with an allocated speaking time of below 5\% throughout both phases.

\begin{figure}[H]
    \centering
    \includegraphics[width=\textwidth]{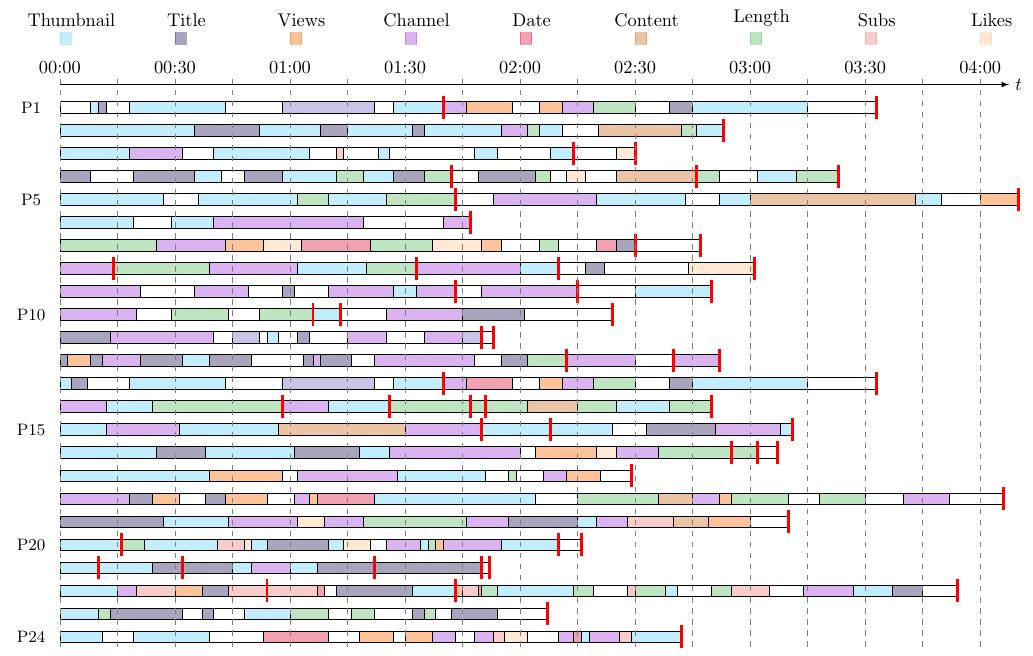}
    \caption{Topography of the think-aloud interviews in the first phase, one for each study participant P1 to P24. The red strokes indicate a decision for or against a video. The upper row indicates the color coding for each surface feature. White sections represent small breaks where the participants did not address a specific surface feature.}
    \label{fig:phase-1} 
\end{figure}

\begin{figure}[H]
    \centering
    \includegraphics[width=\textwidth]{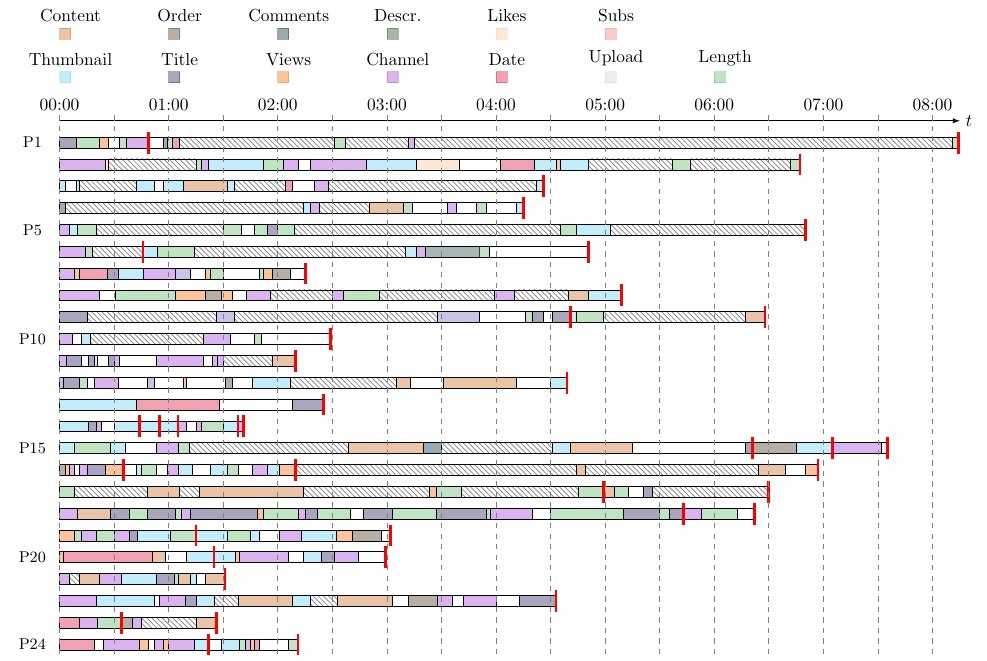}
    \caption{Topography of the think-aloud interviews in the second phase, one for each study participant P1 to P24. The red strokes indicate a decision for or against a video. The upper row indicates the color coding for each surface feature. Hatched sections represent parts of the interview where the participants watched a video.}
    \label{fig:phase-2} 
\end{figure}

A more comprehensive insight into the structure of the participants' selection process is offered by the bar charts in Figures~\ref{fig:phase-1} and~\ref{fig:phase-2}. 

Here, each individual interview is presented as a bar and the sections dedicated to the different surface features are color coded respectively. Hence, these visualizations allow for a deeper insight into the temporal topography of each interview. Analyzing this topography, it becomes apparent that blue (thumbnail) and violet (channel) cover the most area during the first phase, in accordance with the findings presented in Figure~\ref{fig:perc-talk}. In the second phase, where participants were allowed to click on and even watch videos, this dynamic changes: On the one hand, the violet sections increase, indicating a greater focus on the surface feature channel. On the other hand, the participants had access to more surface features such as comments or the video description. Decisions for or against a video are indicated by red strokes. For example, a red stroke after a blue section displays a decision against a video because of the thumbnail. A summary of all decisions and the surface features they were based on is presented in Table~\ref{tab:decisions}.

\begin{table}[h]
\begin{center}
%\begin{minipage}{\textwidth}
\caption{Overview of the decisions for or against a video based on the respective (surface) features, sorted by study phases. }\label{tab:decisions}
\begin{tabular}{p{.23\textwidth}|cc|cc|p{.23\textwidth}}
Surface Feature & \multicolumn{2}{p{.23\textwidth}|}{Decision for (+) and against ($-$) a video in phase 1} & \multicolumn{2}{p{.23\textwidth}|}{Decision for (+) and against ($-$) a video in phase 2} & Total decisions for both phases combined\tabularnewline
&+&$-$&+&$-$&\tabularnewline
\hline 
\hline 
Thumbnail & 3 & 12 & 2 & 9 & 26 (5+, 21$-$)\tabularnewline
%\hline 
Length & 3 & 12 & 2 & 6 & 23 (5+, 18$-$)\tabularnewline
%\hline 
Channel & 3 & 4 & 0 & 2 & 9 (3+, 6$-$)\tabularnewline
%\hline 
Title & 1 & 4 & 1 & 3 & 9 (2+, 7$-$)\tabularnewline
%\hline 
Views & 1 & 1 & 1 & 3 & 6 (2+, 4$-$)\tabularnewline
%\hline 
Content & 0 & 1 & 2 & 0 & 3 (2+, 1$-$)\tabularnewline
%\hline 
Likes & 1 & 1 & 0 & 0 & 2 (1+, 1$-$)\tabularnewline
%\hline 
Subs & 0 & 1 & 0 & 0 & 1 (0+, 1$-$)\tabularnewline
%\hline 
Description & 0 & 0 & 1 & 0 & 1(1+, 0$-$)\tabularnewline
%\hline 
Order & 0 & 0 & 0 & 0 & 0\tabularnewline
%\hline 
Date & 0 & 0 & 0 & 0 & 0\tabularnewline
%\hline 
Comments & 0 & 0 & 0 & 0 & 0\tabularnewline
\hline 
Not related to a specific surface feature & 13 & 1 & 17 & 0 & 31 (30+, 1$-$)\tabularnewline
\end{tabular}
%\end{minipage}
\end{center}
\end{table}

With a total of 26 and 23 decisions, the surface features thumbnail and video length are by far the most influential ones. In congruence to the bar chart presented in Figure~\ref{fig:perc-talk}, the channel, the video title and the number of views views can also be regarded guiding for the selection process, while more specific features such as likes, subs or comments seem almost entirely irrelevant for decision-making. Lastly, it is noticeable, that the most decisions could not be attributed to a specific surface features. In particular, 30 out of the 51 positive decisions do not seem to be related to surface features provided by YouTube. We will elaborate on this finding in more detail in the discussion section~\ref{sec:discussion}.

\begin{figure}[H]
    \centering
    \includegraphics[width=.6\textwidth]{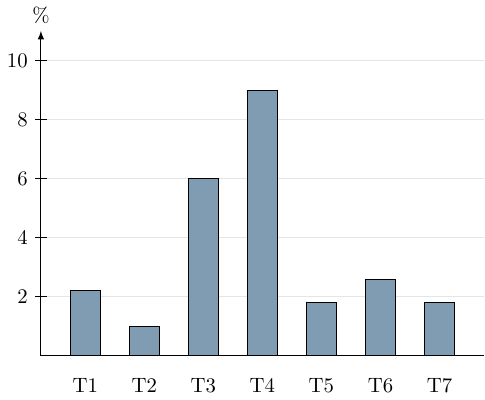}
    \caption{Bar charts visualizing the relative share of each category (T1 to T5) addressing the thumbnail that was used as an argument for or against a video. The categories are summarized in the category system provided in the appendix of this article (cf. Table~\ref{tab:cat-system}).}
    \label{fig:qual-local} 
\end{figure}

To obtain a more in-depth view on the decisions used by participants that could be related to a surface feature, their arguments during the interviews were categorized (cf. Section~\ref{sec:methods}). Since the thumbnail feature lead to the most decisions in both phases, we present a bar chart for the most frequently used arguments in Figure~\ref{fig:qual-local}.

\begin{figure}[H]
    \centering
    \includegraphics[width=.6\textwidth]{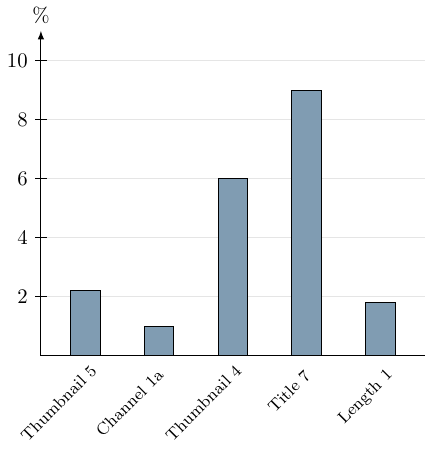}
    \caption{Top five of the most frequently used arguments for or against a video during the think-aloud interviews. The respective category system is provided in the appendix of this article (cf. Table~\ref{tab:cat-system}).}
    \label{fig:top5} 
\end{figure}

With the most frequently used arguments being T3 (the thumbnail indicates an interesting video) and T4 (the thumbnail indicates a boring or nonprofessional video), it becomes apparent that arguments addressing affective aspects dominate over content-related reasonings. An overview of the overall top 5 most frequently used arguments is provided in Figure~\ref{fig:top5}. We discuss these findings in section~\ref{sec:discussion}.

\subsection{Questionnaire results}\label{sec:results-3}
\begin{figure}[H]
    \centering
    \includegraphics[width=\textwidth]{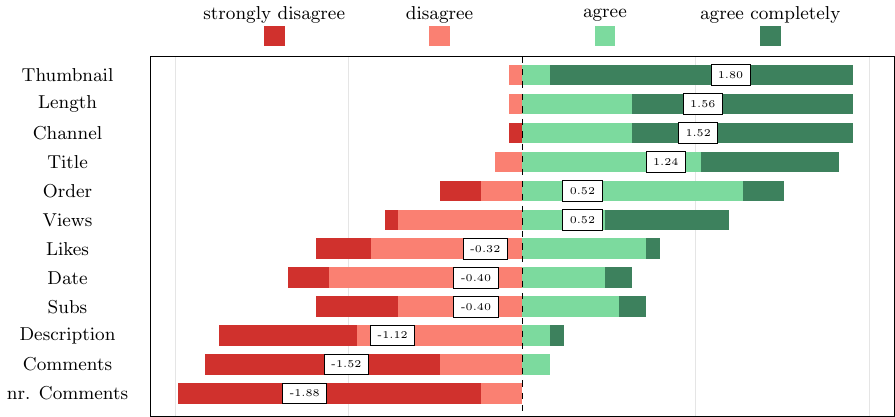}
    \caption{Diverging stacked bar chart visualizing the participants' agreement that the respective surface feature is important for decision making. The respective average ratings for each (surface) feature (cf. Section~\ref{sec:methods}) are provided in labels on each bar where 2 corresponds to ``agree completely'' and $-2$ to ``strongly disagree''. The abbreviation ``nr. Comments'' stands for the number of comments under a video. }
    \label{fig:dsbc} 
\end{figure}

In the final part of our study we, in retrospection, asked the participants to evaluate the extent to which they agree with the respective features having been relevant for their selection process. The responses are visualized using a diverging stacked bar chart in Figure~\ref{fig:dsbc}. In congruence to our previous findings, the thumbnail takes a sole first place with a rating of 1.80 (where 2 corresponds to ``agree completely'' and $-2$ to ``strongly disagree''). With 21 out of 24 participants agreeing completely with the thumbnail being important for their selection process, the thumbnail even exceeds the video length in terms of relevance for decision-making in the participants' retrospective views. In contrast, the number of comments (average rating $-1.88$) and comments themselves (average rating $-1.52$) solidify last places and do not seem to contribute meaningfully to the decision-making process.

\section{Discussion}
\label{sec:discussion}

\subsection{The process of selecting instructional videos for physics teaching}
In phase 1 of our study, pre-service physics teachers were given the task of selecting one explanatory video on quantum physics from a set of eight options. Participants were provided with excerpts from the YouTube search results and had access to various metrics such as views, likes, and channel information. The analysis of participants' eye movements revealed a significant emphasis on video thumbnails, with more than one-third of their total fixation time and counts directed towards this area of interest (AOI). Surprisingly, there were no statistically significant differences in mean fixation duration between different AOIs, despite this measure ``is often considered an indicator of cognitive processing demand'' \citep[][p. 10]{hahn2022eye}. This finding, hence, contrasts with the results of \cite{hsieh2011different}, who suggested that viewing content with different information types requires varying cognitive resources.

To gain further insight into the participants' selection processes, we examined their think-aloud data. This analysis, consistent with the eye-tracking data, revealed that during both phase 1 (selection of one out of eight options based on surface features) and phase 2 (free exploration), participants predominantly voiced their considerations in relation to the thumbnail AOI. An in-depth categorization of argumentation structures and decision-making uncovered four key observations:
\begin{enumerate}
    \item The video duration played a significant role in participants' choices, aligning with didactic perspectives as they were selecting videos for instructional purposes.
    \item The video content had a minor influence on participants' decisions during the free exploration phase. Instead, choices were primarily guided by thumbnail, duration, channel, and title features, indicating a reliance on surface features and pragmatic decision-making among pre-service physics teachers. This tendency to select videos they already had a connection with or were familiar with, such as those from known channels, is in accordance with findings from cognitive psychology \citep{chen2016things}.
    \item A notable portion of the decisions made during the study could not be attributed to surface features, comments, or video content based on either eye-movements or verbalizations. Hence, in these cases, the study participants either struggled to explicitly articulate their decisions due to multiple considerations or did just not articulate them at a deeper level. Similar cases have been observed in physics education research on teachers' professional  competences, where prior research has found that experienced teachers' actions in the classroom are guided by informed decisions and teaching routines that cannot be easily verbalized \citep[e.g., see][]{borko1989cognition,livingston1989expert,stender2014unterrichtsplanung}. To clarify whether similar principles contribute to an explanation of the observations made in our study requires further investigation.
    \item Although the participants had the opportunity to view comments associated with each YouTube video during the free exploration phase 2, surprisingly, none of the participants explicitly based their decisions on comments. This observation contrasts with findings from previous research reporting that students relied on comments as an indicator of video quality \citep[e.g., see][]{fyfield2021navigating,tan2011open}  and, instead, indicates that the selection process tends to be less systematic. However, previous studies have consistently shown a strong and statistically significant correlation between the explanatory quality of YouTube videos and the number of content-related comments \citep{kulgemeyer2016,bitzenbauer2023}. It is therefore noteworthy that these comments did not play a significant role in the decision-making process of our participants.
\end{enumerate}
The eye-tracking and think-aloud data were complemented by retrospective questionnaire responses: When asked about the features that influenced their video selection for instructional purposes, the majority of respondents indicated thumbnail, duration, channel, or title, while video descriptions and the quantity or quality of comments played a minor role, even in retrospective evaluation.

The finding that participants primarily explored the top results of the YouTube search list aligns with previous studies \citep[e.g., see][]{fyfield2021navigating,tan2011open}: Over half of the participants reported approaching video selection in a sequential manner based on the order of videos in the search list (cf. figure~\ref{fig:dsbc}). This pragmatic approach leads to quick decisions (made in a time frame of less than 10 minutes in the free exploration phase 2 of this study) that are mainly based on surface features (e.g., thumbnails) or familiarity (e.g., channel). The analysis of individual argumentation categories (see figure~\ref{fig:qual-local} and figure~\ref{fig:top5})  supports this assumption.

In conclusion, our findings suggest that the decision-making process of (pre-service) physics teachers when searching for suitable YouTube explanatory videos (on quantum topics in this study) for instructional purposes is primarily driven by pragmatism, efficiency, and reliance on familiar features. The  availability of empirical evidence regarding the explanatory quality of online videos seems to be  overlooked by (pre-service) physics teachers, representing a missed opportunity to streamline the selection process. Therefore, in section~\ref{sec:implications}, we will synthesize existing empirical evidence and propose a preliminary decision tree that may assist teachers in efficiently identifying high-quality explanatory videos on YouTube.

\subsection{Contrasting the selection processes with the proposed decision tree}

The observations made in this study indicate that the selection processes of (pre-service) physics teachers when searching explanatory videos suitable for physics teaching are predominantly unsystematic, relying on superficial or familiar aspects, and characterized by pragmatic choices. These tendencies give rise to two interconnected issues when it comes to real instructional preparation, where videos with high explaining quality are sought:

\begin{enumerate}
    \item Teachers may require significant time to find suitable videos due to the unsystematic approach.
    \item There is a probability that teachers may select videos of lower quality.
\end{enumerate}

It is obvious that in in the study reported in this article the decision-making process of the participants in most cases diverged from the proposed decision tree. In light of this, it becomes necessary to support teachers in systematizing their selection process to overcome the identified problems in practice. The decision tree proposed in section~\ref{sec:RQ} might be a valuable tool in this regard as it reflects the sate of the literature. While we are aware that future studies are required and might lead to a refined version of the decision tree, the significance of the decision tree in its current form lies in its capacity to systematize the selection process without imposing quantitative guidelines or thresholds. This acknowledges the absence of empirical evidence supporting such measures and recognizes that decisions should be made by teachers on a case-by-case basis, taking into account the specific topic. Future studies should investigate whether the use of the decision tree indeed facilitates efficient and successful identification of high-quality explanatory videos on various topics. Additionally, it will be crucial to determine whether decision steps need to be supplemented or specified.

\subsection{Limitations}
The present study has several limitations that should be considered when interpreting the results. First, the focus on explanatory videos related to three specific quantum topics may restrict the generalizability of the findings. Although we designed the study with three separate groups, each tasked with selecting videos for instructional situations on two different quantum topics, this control measure may not account for potential variations that could arise if explanatory videos on further (physics) topics, e.g., classical mechanics topics, were included. Further research is needed to validate the reported results in a broader range of topics. Second, understanding the cognitive processes of prospective physics teachers during video selection is a complex empirical endeavor, and the chosen data collection methods -- even though they might complement each other -- come with inherent limitations. While the analysis of eye-tracking data is based on the eye-mind assumption \citep{just1980theory}, previous research has emphasized the importance of complementing eye movement analysis with additional verbal data to gain a comprehensive understanding \citep{bruckner2020epistemic,chien2015learning,chiou2022exploring,mason2013eye,smith2010eye,wu2021eye}. To address this concern within our mixed-methods approach, we employed introspective thinking-aloud in our study. Additionally, the retrospective questionnaire used for internal validation allows participants to reflect on their experiences; however, it may also trigger ad-hoc generated associations and thoughts regarding the different YouTube surface features \citep[for similar arguments see][]{winkler2021quantum,winkler2023european}. Third, it is important to consider that while this study focuses on the selection processes employed by (pre-service) physics teachers in finding YouTube explanatory videos on quantum physics suitable for teaching, the selection situations created within the study design differ from real classroom planning scenarios. Particularly, in our study, participants had unlimited time for decision-making, whereas real instructional planning is significantly influenced by time constraints. However, the analysis of think-aloud data reveals that decision-making occurred within a time frame of less than 10 minutes in phase 2 of the study (free exploration), which aligns with a reasonable time frame in natural classroom planning situations. Lastly, the time-consuming nature of the study led to a relatively small sample size. However, the primary aim of this study was to gain in-depth insights into the video selection process rather than to produce generalizable findings on a surface level. Future research with larger sample sizes could provide a broader perspective on the topic.

\section{Conclusion}
\label{sec:implications}
This mixed-methods study explored how pre-service teachers select instructional videos on YouTube for physics teaching, focusing on the role of surface features (likes, views, thumbnails) and comments. The results indicate that the decision-making processes of (pre-service) physics teachers when searching for suitable YouTube explanatory videos is primarily driven by pragmatism, efficiency, and reliance on familiar features. 

Based on the current sate of research into the explaining quality of online explanatory videos, we proposed a decision tree which reflects how an efficient and successful selection process might look. Although the decision-making process of the study participants often differed from the proposed decision tree in this study, it serves as a hypothesis for future research aimed at supporting teachers in systematizing their selection process: Further studies should explore whether the decision tree facilitates efficient and successful identification of high-quality explanatory videos on various topics, and how it might be adapted and refined, e.g., with regards to different subject areas and teaching contexts. Also, future studies might examine how the decision tree works as a tool for preparing teaching, e.g., in related courses in science teacher education. Lastly, it seems particularly crucial to consider the evolving nature of online platforms in future research: For example, research could examine how the decision tree (or an evolved version thereof) can be adapted to accommodate changes in platform features and the emergence of new video metrics. Collaborative research involving educators, researchers, and platform developers may further enhance the decision tree's practicality and usability for (pre-service) teachers, facilitating their video selection process and ultimately benefiting student learning experiences in physics and beyond.

\backmatter

% \bmhead{Supplementary information}

% If your article has accompanying supplementary file/s please state so here. 

% Authors reporting data from electrophoretic gels and blots should supply the full unprocessed scans for key as part of their Supplementary information. This may be requested by the editorial team/s if it is missing.

% Please refer to Journal-level guidance for any specific requirements.

% \bmhead{Acknowledgments}

% Acknowledgments are not compulsory. Where included they should be brief. Grant or contribution numbers may be acknowledged.

% Please refer to Journal-level guidance for any specific requirements.

\section*{Declarations}
\bmhead{Funding} 
No funding was received for conducting this study.

\bmhead{Competing interests}
The authors have no competing interests to declare that are relevant to the content of this article.

\bmhead{Financial Interests}
The authors have no relevant financial or non-financial interests to disclose.

\bmhead{Ethics approval}
Ethical review and approval was not required for the study on human participants in accordance with the local legislation and institutional requirements. The patients/participants provided their written informed consent to participate in this study.

\bmhead{Informed consent of participants}
Not applicable.

\bmhead{Data availability}
The data presented in this study are available on request from the corresponding author.

% \begin{itemize}
% \item Funding
% \item Conflict of interest/Competing interests (check journal-specific guidelines for which heading to use)
% \item Ethics approval 
% \item Consent to participate
% \item Consent for publication
% \item Availability of data and materials
% \item Code availability 
% \item Authors' contributions
% \end{itemize}

% \noindent
% If any of the sections are not relevant to your manuscript, please include the heading and write `Not applicable' for that section. 

%%===================================================%%
%% For presentation purpose, we have included        %%
%% \bigskip command. please ignore this.             %%
%%===================================================%%
% \bigskip
% \begin{flushleft}%
% Editorial Policies for:

% \bigskip\noindent
% Springer journals and proceedings: \url{https://www.springer.com/gp/editorial-policies}

% \bigskip\noindent
% Nature Portfolio journals: \url{https://www.nature.com/nature-research/editorial-policies}

% \bigskip\noindent
% \textit{Scientific Reports}: \url{https://www.nature.com/srep/journal-policies/editorial-policies}

% \bigskip\noindent
% BMC journals: \url{https://www.biomedcentral.com/getpublished/editorial-policies}
% \end{flushleft}

%%===========================================================================================%%
%% If you are submitting to one of the Nature Portfolio journals, using the eJP submission   %%
%% system, please include the references within the manuscript file itself. You may do this  %%
%% by copying the reference list from your .bbl file, paste it into the main manuscript .tex %%
%% file, and delete the associated \verb+\bibliography+ commands.                            %%
%%===========================================================================================%%

\bibliography{sn-bibliography}% common bib file
%% if required, the content of .bbl file can be included here once bbl is generated
%%\input sn-article.bbl

%% Default %%
%%\input sn-sample-bib.tex%

\section*{Appendix}\label{sec:appendix}

\begin{table}[h]
\begin{center}
\caption{Category system used for the think aloud study.}\label{tab:cat-system}
\begin{tabular}{c|p{10cm}}
\hline 
Surface Feature & Category\tabularnewline
\hline 
\hline 
\multirow{8}{*}{Thumbnail} & 1. The thumbnail indicates a thematically well-suited video\tabularnewline
 & 2. The thumbnail indicates that the video might be off-topic\tabularnewline
 & 3. The thumbnail indicates that the video might not be exhaustive,
e.g. it is part of a longer series\tabularnewline
 & 4. The thumbnail seems interesting \tabularnewline
 & 5. The thumbnail seems unappealing or non-professionell\tabularnewline
 & 6. The thumbnail indicates that the topic is explained well\tabularnewline
 & 7. The thumbnail indicates that the topic is not explained well\tabularnewline
 & 8. Neutral examination of the thumbnail\tabularnewline
\hline 
\multirow{7}{*}{Title} & 1. The title indicates a thematically well-suited video\tabularnewline
 & 2. The title indicates that the video might be off-topic\tabularnewline
 & 3. The thumbnail indicates that the video might not be exhaustive,
e.g. it is part of a longer series\tabularnewline
 & 4. The title seems interesting\tabularnewline
 & 5. The title seems unappealing or non-professionell\tabularnewline
 & 6. The title indicates that the video might delve too deeply into
the topic\tabularnewline
 & 7. Neutral examination of the title\tabularnewline
\hline 
\multirow{4}{*}{Subscriptions} & 1. The respective channel has many subscriptions\tabularnewline
 & 2. The respective channel has few subscriptions\tabularnewline
 & 3. The respective channel has more subscription than other channels\tabularnewline
 & 4. The respective channel has fewer subscriptions than other channels\tabularnewline
\hline 
\multirow{10}{*}{Channel} & 1. The channel is known\tabularnewline
 & (a) The channel's name is known\tabularnewline
 & (b) The channel is known for high-quality videos\tabularnewline
 & (c) The channel is known for scientifically inaccurate videos\tabularnewline
 & (d) The channel is known for low-quality videos\tabularnewline
 & 2. The channel is unknown\tabularnewline
 & 3. Neutral examination of the channel\tabularnewline
 & 4. The channel seems fitting\tabularnewline
 & 5. The channel seems unfitting\tabularnewline
 & 6. The channel is verified (blue checkmark)\tabularnewline
\hline 
\multirow{4}{*}{Likes} & 1. The video has many likes\tabularnewline
 & 2. The video has few likes\tabularnewline
 & 3. The video has more likes than other videos\tabularnewline
 & 4. The video has fewer likes than other videos\tabularnewline
\hline 
\multirow{5}{*}{Views} & 1. The video has many views\tabularnewline
 & 2. The video has few views\tabularnewline
 & 3. The video has more views than other videos\tabularnewline
 & 4. The video has fewer views than other videos\tabularnewline
 & 5. Neutral examination of views\tabularnewline
\hline 
\multirow{5}{*}{Length} & 1. The video length seems too long\tabularnewline
 & 2. The video length seems accurate\tabularnewline
 & 3. The video length seems rather short\tabularnewline
 & 4. The video length seems too short\tabularnewline
 & 5. Netrual examination of video length\tabularnewline
\hline 
\multirow{3}{*}{Upload} & 1. The video is too old\tabularnewline
 & 2. The video is new\tabularnewline
 & 3. Neutral examination of upload date\tabularnewline
\hline 
\multirow{2}{*}{Order} & 1. The videos suggested first are fitting\tabularnewline
 & 2. The videos suggested first are unfitting\tabularnewline
\hline 
\multirow{2}{*}{Description} & 1. Neutral examination of video description\tabularnewline
 & 2. The video description is appeealing\tabularnewline
\hline 
\multirow{4}{*}{Content} & 1. The video content could be fitting\tabularnewline
 & 2. The video content seems rather unfitting\tabularnewline
 & 3. Neutral examination of video content\tabularnewline
 & 4. The video content is contingent\tabularnewline
\hline 
\multirow{2}{*}{Comments} & 1. Checking the feedback emerging from comments\tabularnewline
 & 2. The comments are rather positive\tabularnewline
\hline 
\end{tabular}
\end{center}
\end{table}

\end{document}